\documentclass[twoside,twocolumn]{article}

\usepackage{amsmath}
\usepackage{abstract}
\usepackage{titletoc}
\usepackage{titlesec}
\usepackage[normalem]{ulem}

\usepackage{tocloft}
\usepackage[utf8]{inputenc}
\usepackage{graphicx}
\usepackage{multicol}

\graphicspath{ {figures/} }
\usepackage[backend=biber,style=ieee,sorting=none]{biblatex}
\usepackage[font=footnotesize]{caption}
\usepackage{gensymb}
\usepackage{subcaption}
\usepackage{tabularx}
\usepackage{tabulary}
\usepackage{lscape} 
\usepackage{hhline}
\usepackage[symbol]{footmisc}
\usepackage{booktabs}
\usepackage{multirow}
\usepackage{floatrow}
\usepackage{float}
\usepackage{color}
\usepackage{colortbl}
\usepackage[table,dvipsnames]{xcolor}
\definecolor{Gray}{gray}{0.9}
\definecolor{AB}{rgb}{0.8, 0.9, 1.0}
\usepackage[a4paper, margin=.5In, bottom=1.3In, bindingoffset=0mm]{geometry}
\usepackage{setspace}
\usepackage{txfonts}
\usepackage[nottoc]{tocbibind}

\usepackage{abstract} 

\DeclareUnicodeCharacter{2060}{\nolinebreak}
\usepackage{titling} 

\usepackage{hyperref} 

\pretitle{\begin{center}\Huge\bfseries} 
\posttitle{\end{center}} 
\title{Neutron Production in The Scattering S250 Mevion Proton-Therapy System} 
\author{%
{Nima Tatari}\\[1ex] 
\normalsize Washington University in St. Louis, Department of Physics \\ 
\normalsize \href{mailto:n.tatari@wustl.edu}{n.tatari@wustl.edu}
}
\date{\today} 
\renewcommand{\maketitlehookd}{%
\begin{abstract}

\textbf{Purpose:} Secondary radiation in a radiation therapy environment is considered a cause for post-treatment side effects including secondary carcinogenesis. In a proton-therapy setup, neutrons deliver a considerable extra dose to the patient. Various measurements and calculations of neutron dose suggest a wide range of values for this quantity. Based on work with a Mevion S250 passive scattering proton-therapy system, this paper provides a comparison between the previously published values and analyzes the Monte Carlo calculated neutron dose from the present work using the neutron cross-section data. The purpose is to provide a range for the convergence of several neutron-absorbed dose data points (e.g. 3.94 mSv/Gy in a Bonner sphere measurement and 0.4 mSv/Gy in a SWENDI measurement) in previous works of other authors.

\textbf{Methods:} The analysis is based on the neutron spectrum in the regions of thermal to evaporation energies. Data for the neutrons is obtained by the Geant4 Monte Carlo simulation toolkit for particle transport. A water phantom of dimensions $35 \times 35 \times 35 cm^{3}$ is positioned at the center of the treatment room (i.e. the isocenter) modeling the human’s body. The calculated spectra are compared to the previous measurements. Tables of neutron cross-sections are provided to explain the spectral behavior of the neutrons.

\textbf{Results:} Several items in the room contribute differently to the dose absorbed by the human body. These are scored separately and presented in pie charts. The calculated total amount of the absorbed neutron dose values are provided along and perpendicular to the beamline, ranging from 1 to 20 mSv/Gy depending on how close to the neutron source the data is obtained.

\textbf{Conclusions:} Matching the calculated neutron dose data with the previously measured values within a range below $1\%$ provides further evidence for the accuracy of the measurement method, i.e. extended range Bonner sphere (ERBS). The percentage contribution of the internally produced neutrons to the total absorbed dose was studied by varying the beam energy. We concluded that the reducible part of the neutron dose absorbed to the body, i.e. the externally produced neutrons, comprises less than $50 \%$ of the total.

\textbf{Keywords:} Proton-therapy, Mevion, Secondary radiation, Neutron absorbed dose.
\end{abstract}

}

\begin{document}

\maketitle	


\section{Introduction}
A study of the neutron production in a single-room Mevion S250 passive scattering proton-therapy system is performed. The proton-therapy system has been in operation at Washington University in St. Louis since 2013 and was decommissioned in 2020. Geant4 Monte Carlo toolkit has been used to calculate neutron fluence spectra from various nozzle components. Each fluence spectrum is discussed based on the cross-sections of the materials contributing to secondary neutron production.

\section{Materials and methods}
Several parameters in the simulation consist of the geometrical dimensions of the room, synchrocyclotron, and nozzle components. The nozzle includes a first scatterer, a range modulation wheel, the main range shifter, a second scatterer, a beam collimation system, a brass aperture, and a range compensator. A water phantom models the patient’s body. The geometry and physics models in the simulation are discussed in the following section.

\subsection{Geometry}
Modeling a geometry in Geant4 is based on the dimensions, compositions, and densities of the objects in the environment of radiation transport. The treatment room, synchrocyclotron, beam-delivery system, and water phantom define the geometry, each part contributing to the total neutron production based on different physical processes. The neutrons from the room are created by previously generated neutrons in the nozzle. The contribution of the protons that are lost in the synchrocyclotron's extraction stage to the production of secondary neutrons is calculated. Figures 1 (a,b,c) show the geometry of the room containing the proton-therapy system. The whole vault is modeled with a 13.5$\times$15 m$^2$ floor and with a 15 m height.

\begin{figure}[ht!]
\centering
\includegraphics[scale=0.09]{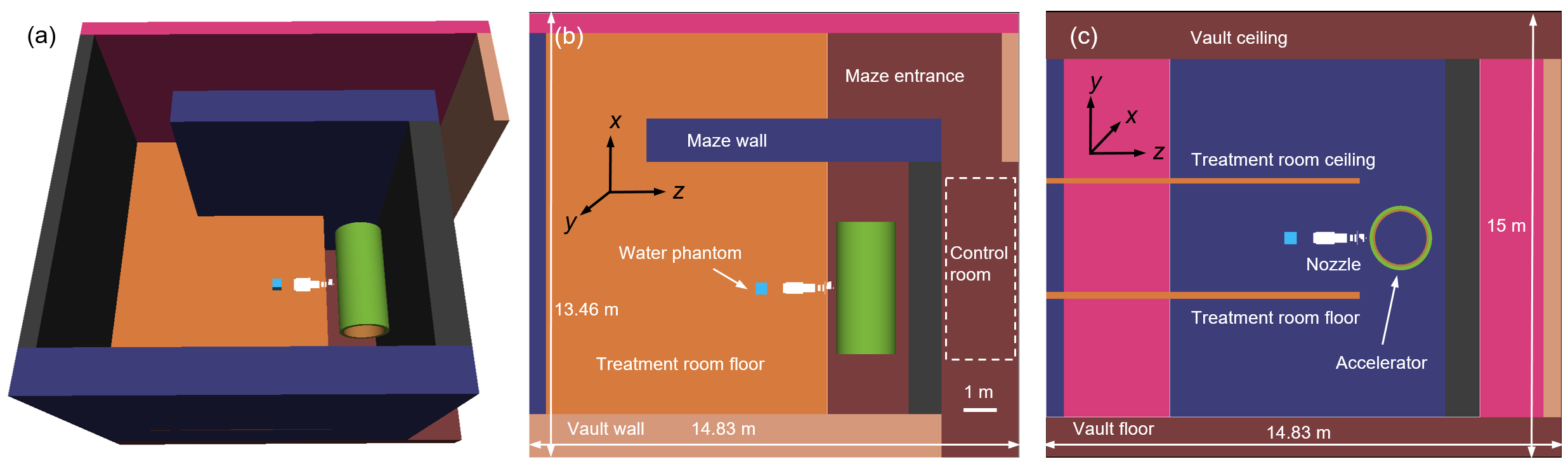}
\caption{a) 3D view, b) top view, and c) side view of the treatment room geometry design in the Geant4 model of the Mevion S250 scattering proton system.}
\end{figure}

The components of the nozzle are shown in Figure 2. The first scatterer is positioned after the beam extraction point from the synchrocyclotron.

\begin{figure}[ht!]
\centering
\includegraphics[scale=0.11]{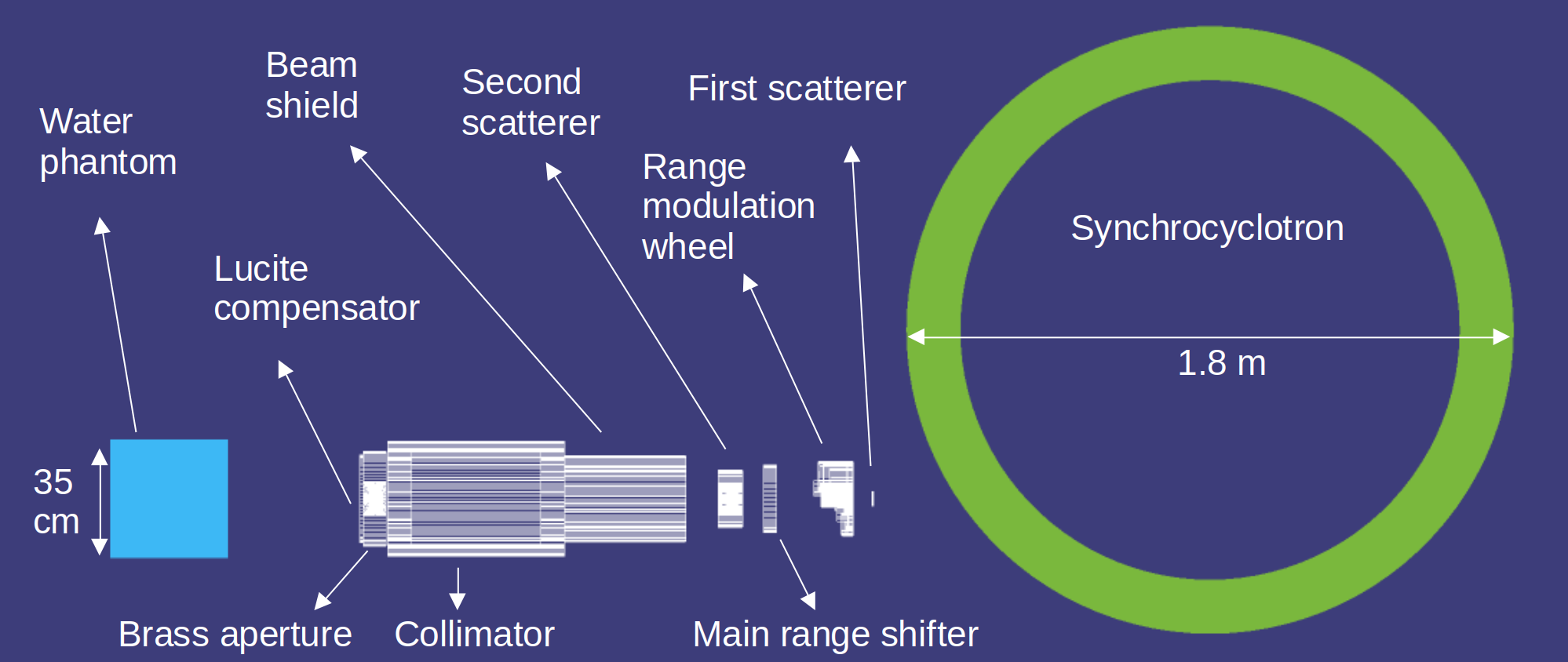}
\caption{Nozzle components in a Geant4 model of Mevion S250 scattering proton system.}
\end{figure}

The scattering proton machine is modeled with a primary mono-energetic proton beam of 250 MeV and is spread laterally to get an even dose throughout a specific area and depth. Rotation of the RMW superposes the modulated proton beams with different energies. The result is a spread-out Bragg peak (SOBP), with a uniform proton dose curve, inside the water phantom.

Broadening of the proton beam in the model is achieved through the second scatterer. The unwanted part of the beam is trimmed by the collimation system which is modeled by highly attenuating materials such as stainless steel and brass with large stopping power. The collimation system is modeled with a stainless-steel cylinder with a 3 cm thickness, 9.75 cm inner radius, and 36 cm height. A similar structure with a thickness of 2 cm, 15 cm inner radius, and 52.5 cm height was used for the collimator with the same material. In addition to the outer cylinder, the collimator contains two inner rings at its entrance and exit sides. Each ring has an inner radius of 12.5 cm with the same material as the outer collimator. A brass aperture with a thickness of 6.84 cm is placed next to the collimator. The center of the brass aperture contains a square opening.

During the actual treatment, the fine-tuning of the output beam energy is done by a patient-specific lucite compensator that varies the depths of the beam at different positions to shield the healthy tissues from the proton beam. The compensator is modeled with a lucite cylinder of 5 cm height placed as the last step of the nozzle.

\subsection{Geant4 Monte Carlo simulation toolkit}
Geant4 (GEometry ANd Tracking) has been developed by CERN as a general-purpose Monte Carlo (MC) simulation software for particle transport through matter. Two fundamental aspects of Geant4 are handling the geometrical shapes and tracking various particles through objects made of different materials. Detailed radiation transport calculations are performed by the toolkit for that purpose. The building blocks of Geant4 simulations are G4Step objects from which information can be stored. Particle transport involves a 3D simulation of the interactions of particles irradiated through materials with specific geometries. In this work, the primary particle source is protons.

\section{Results and discussion}
The proton PDDs and neutron spectra are benchmarked with measured data. The neutron fluence spectra calculated by Geant4 form the basis of the presented analysis. The properties of the spectra are described based on the neutron cross-section data obtained from ENDF datasets \cite{39}⁠.

Protons form a well-defined dose deposition peak as they travel through matter, due to the stopping power. The simulation PDD is obtained by emitting a primary proton beam with 250 MeV energy, 2 MeV energy bandwidth, and 1 mm spot size.

\begin{figure}[ht!]
\centering
\includegraphics[scale=0.21]{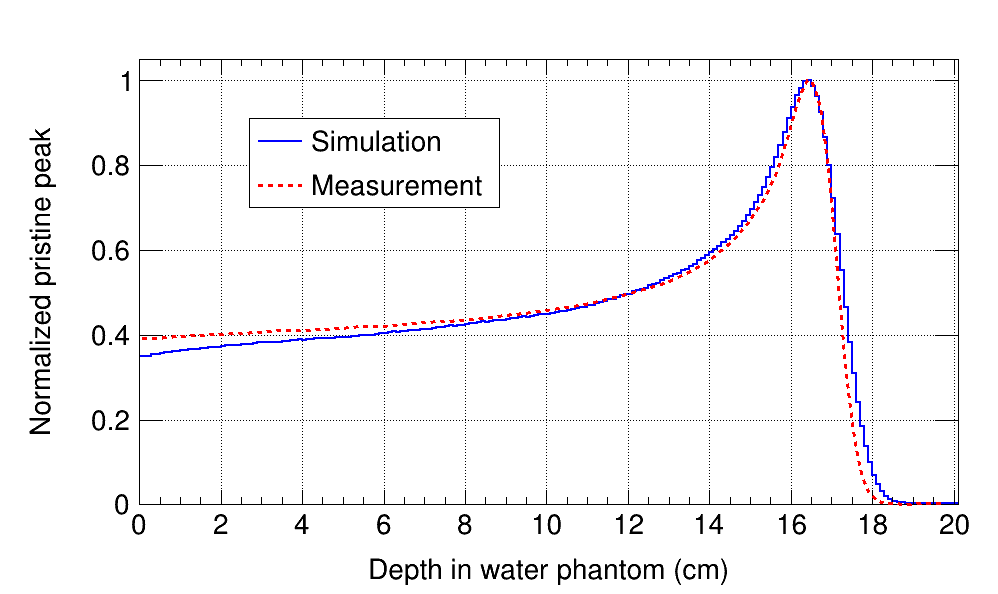}
\caption{Matching the Geant4-calculated pristine percentage depth dose to the experimental data. Both PDDs correspond to the first step of the range modulation wheel. The benchmarked parameters include the thickness of the range shifter (affecting the range in water), the energy width of the primary proton beam (affecting the range straggling), and the spot size of the initial beam (affecting the entrance dose height).}
\end{figure}

Each range modulation step delivers a specific proton beam energy to the isocenter with a corresponding pristine peak depth. Rotation of the range modulation wheel leads to a simultaneous superposition of the proton beams with various energies, and depending on the design of the angular extent of each step that corresponds to their weights, leads to a spread-out Bragg peak or SOBP.

The main characteristics of an SOBP are range and modulation. The range is defined as the distance from the surface of the water phantom to the distal 90\% proton dose. Modulation is defined as the distance from the proximal 95\% dose to the distal 90\% \cite{24}⁠. Specifically, for the treatment setup of the whole brain the range of 17 cm and modulation of 16 cm has been used in the work of Howell et al. \cite{24}. The same SOBP has been simulated in this work for benchmarking the neutron calculations.

Simulation of an SOBP requires a linear combination of several pristine peaks with different ranges. Each pristine peak is obtained by positioning a step of the range modulation wheel in the primary proton beam path. The final SOBP has a 17 cm range and 16 cm modulation. The first 15 pristine peaks delivered by the 15 are used as the base functions in a singular value decomposition (SVD) method to obtain the composition weights. Figures 4 and 5 show the normalized pristine peaks and the SOBP with a range of 17 cm, respectively.

\begin{figure}[ht!]
\centering
\includegraphics[scale=0.21]{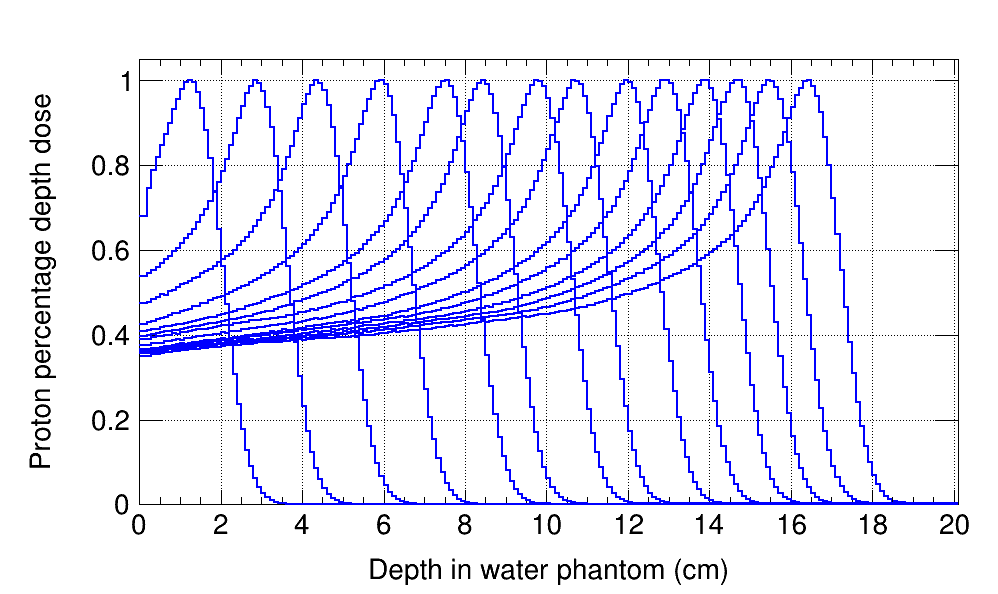}
\caption{Normalized proton pristine peaks corresponding to 15 steps of RMW.}
\end{figure}

\begin{figure}[ht!]
\centering
\includegraphics[scale=0.21]{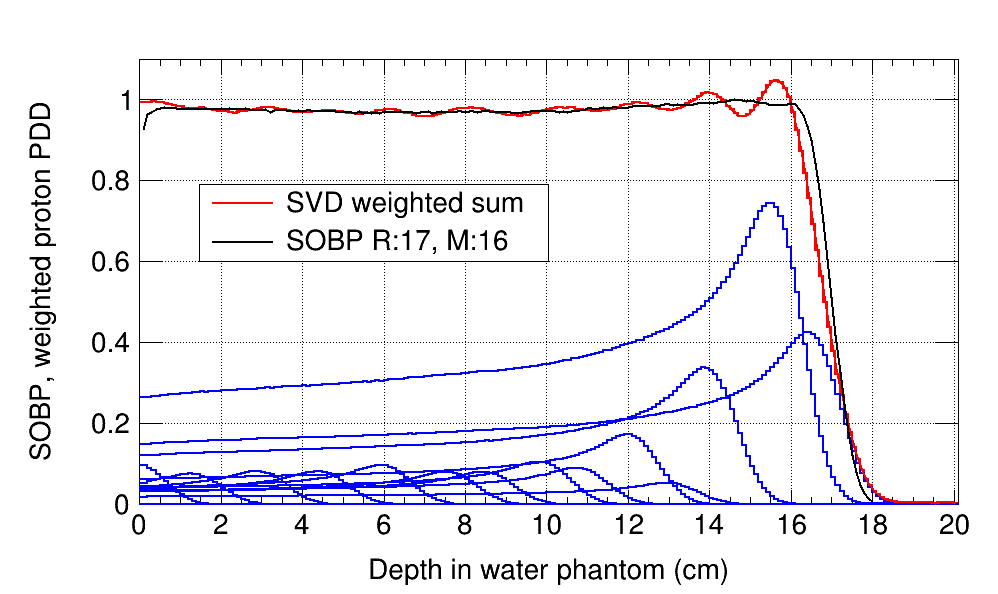}
\caption{Constructing the SOBP from the 14 pristine Bragg peaks using the singular value decomposition method.}
\end{figure}

To obtain the agreement between the transverse proton dose profiles, the parameters that define the shape of the second scatterer are modified. The second scatterer consists of bell-shaped, concave Lexan, and thin, convex lead components. Contracting or widening these parts determines the uniformity of the transverse proton dose profile.

\begin{figure}[ht!]
\centering
\includegraphics[scale=0.21]{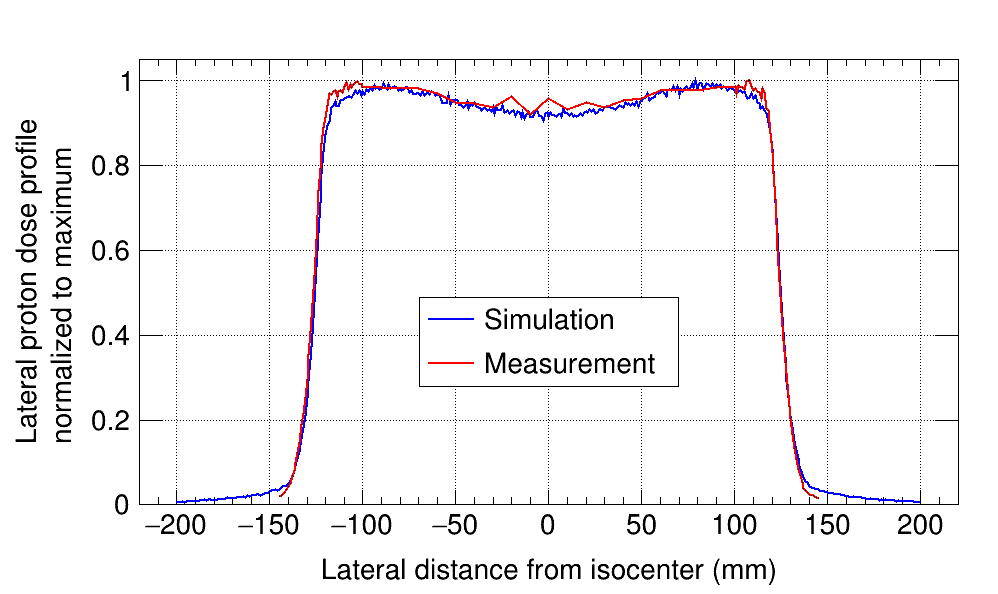}
\caption{Transverse matching of the simulated proton dose profile with the measurement data (Zhao et al. 2016 \cite{9}⁠).}
\end{figure}

In this article, neutron dose-equivalent H* normalized by the treatment dose at the isocenter D is analyzed. The dose D is defined as energy per unit mass, and as a point quantity, it has a specific value at each geometric point. Dose to the isocenter in a passive system per primary proton is in the range of 5.5$\times$10$^{-13}$ Gy/pr to 6.5$\times$10$^{-13}$ Gy/pr. This value remains roughly stable regardless of the field size and delivered proton energy.

\subsubsection{Benchmarking of the neutron fluence and neutron dose spectra}

In a study by Howell et al \cite{24}⁠ \hspace{0.5mm} extended-range Bonner spheres (ERBS) have been experimentally employed and the generated data unfolded to obtain the neutron fluence spectra at 50 cm transverse to the isocenter in a scattering Mevion proton-therapy system. The same geometry is set up with a beam angle in the horizontal plane. Simulations corresponding to different energy outputs are performed.

The neutron fluence spectra, corresponding to each range modulation step, are weighted and summed according to the SOBP weights. Figure 7 indicates the absolute values of the neutron fluence spectra per unit treatment (proton) Gy to the isocenter per unit lethargy. Integration of the curves results in the total fluence per proton Gy. The measured value by the Bonner sphere technique has been reported 1.52$\times$10$^7$ n cm$^{-2}$ Gy$^{-1}$ \cite{24}⁠. The corresponding Geant4 total neutron fluence is 1.45$\times$10$^7$  n cm$^{-2}$ Gy$^{-1}$. The values agree within 4.6\%.

\begin{figure}[ht!]
\centering
\includegraphics[scale=0.21]{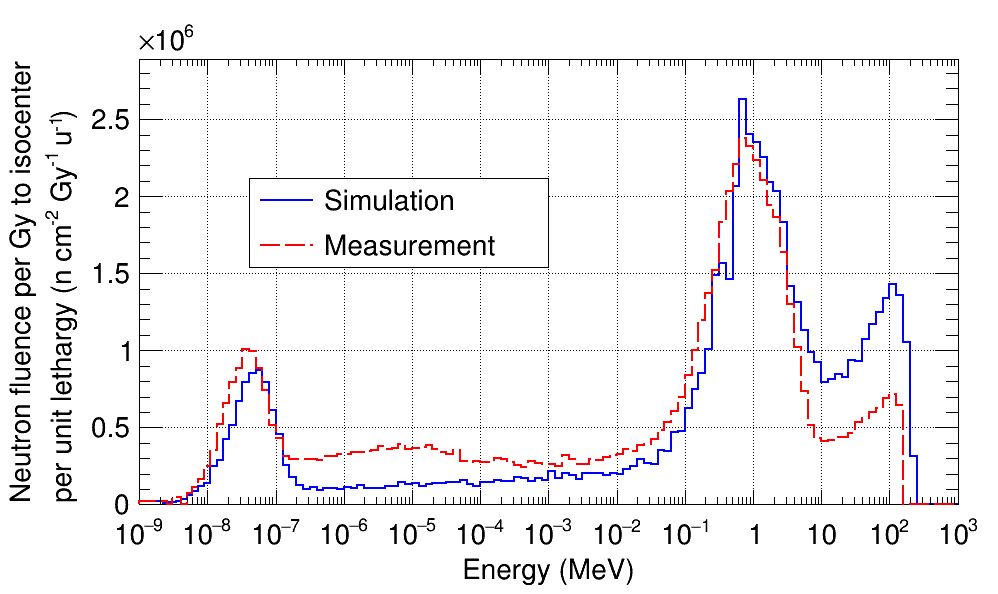}
\caption{Benchmarking of the neutron fluence per treatment proton Gy to the isocenter per unit lethargy. The red curve shows the result of an ERBS measurement by Howell et al. The blue curve shows the Geant4 simulation result of the present work.}
\end{figure}

\begin{figure}[ht!]
\centering
\includegraphics[scale=0.21]{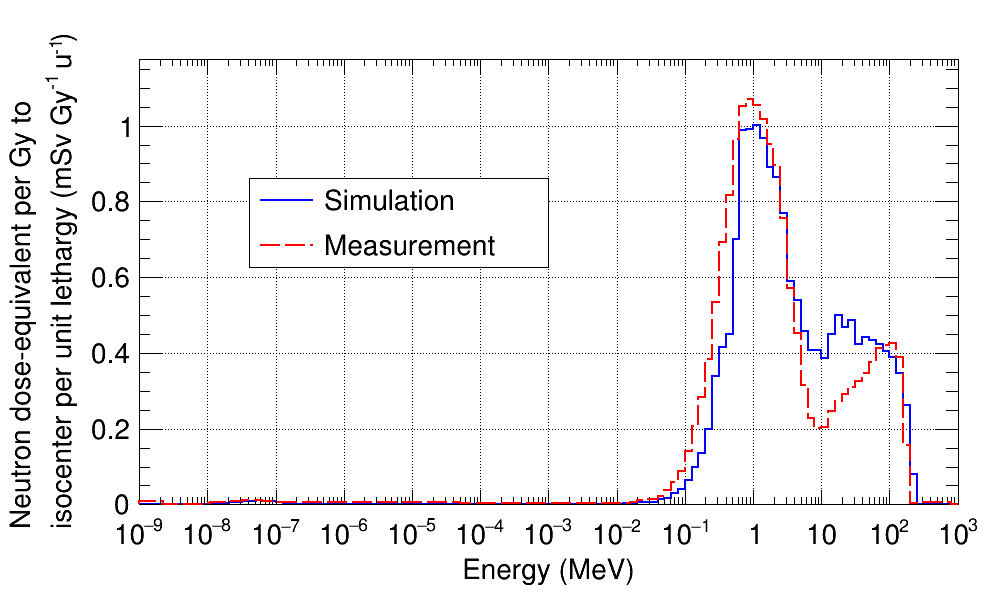}
\label {fig:2_8}
\caption{Benchmarking of the neutron dose-equivalent per treatment proton Gy to the isocenter per unit lethargy. The red curve shows the result of an ERBS measurement by Howell et al. The blue curve shows the Geant4 simulation result of the present work.}
\end{figure}

\subsection{Analysis of the neutron sources using fluence ($\phi$(E)/D)}
Characterizing the source of the neutrons incident onto the water phantom is performed by decomposing the neutron fluence spectrum into several contributions. The cross-section data are used to describe the behavior of the spectra. The synchrocyclotron is made of borated polyethylene and stainless-steel with radii of 74 to 80 cm and 80 to 90 cm respectively and 4 m height is positioned horizontally behind the nozzle.

To account for the proton extraction step inside the synchrocyclotron, a cone region with 90\degree \hspace{0.5mm} opening is considered in its interior part. According to Figure 9 (a), the neutrons produced in the cyclotron contain a thermal-to-evaporation peak ratio close to 1. The borated-polyethylene layer has a large thermalizing effect on the neutrons. The relatively large continuum of the neutrons between the evaporation and thermal regions could be explained based on the multiple objects that a neutron might interact with along its trajectory.

\begin{figure}[ht!]
\begin{subfigure}{\textwidth}
\centering
\includegraphics[scale=0.21]{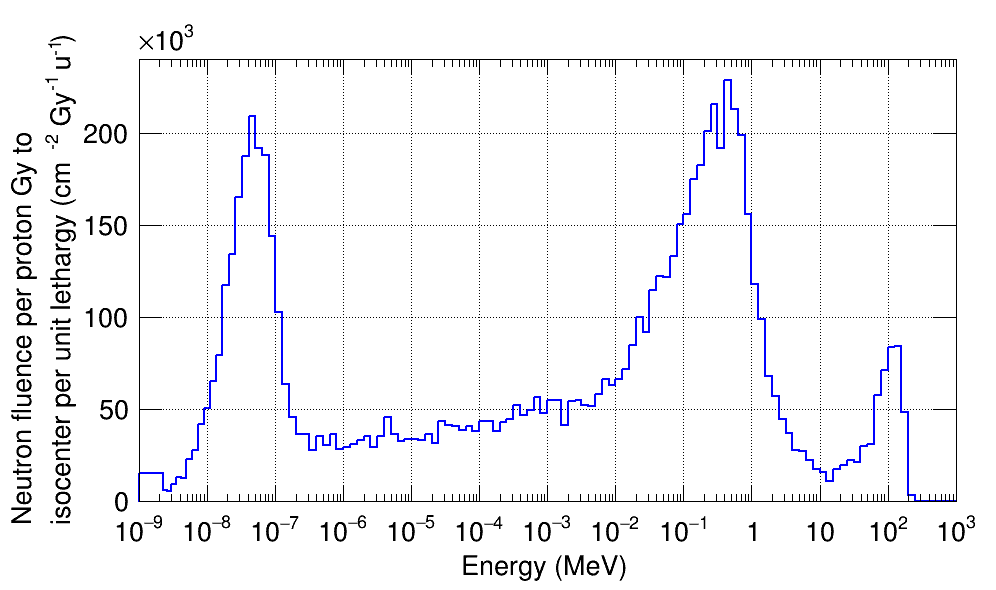}
\caption{Neutrons from the primary protons lost inside the synchrocyclotron}
\label {fig:2_9_a}
\end{subfigure}
\end{figure}
\begin{figure}[ht!]\ContinuedFloat
\begin{subfigure}{\textwidth}
\centering
\includegraphics[scale=0.21]{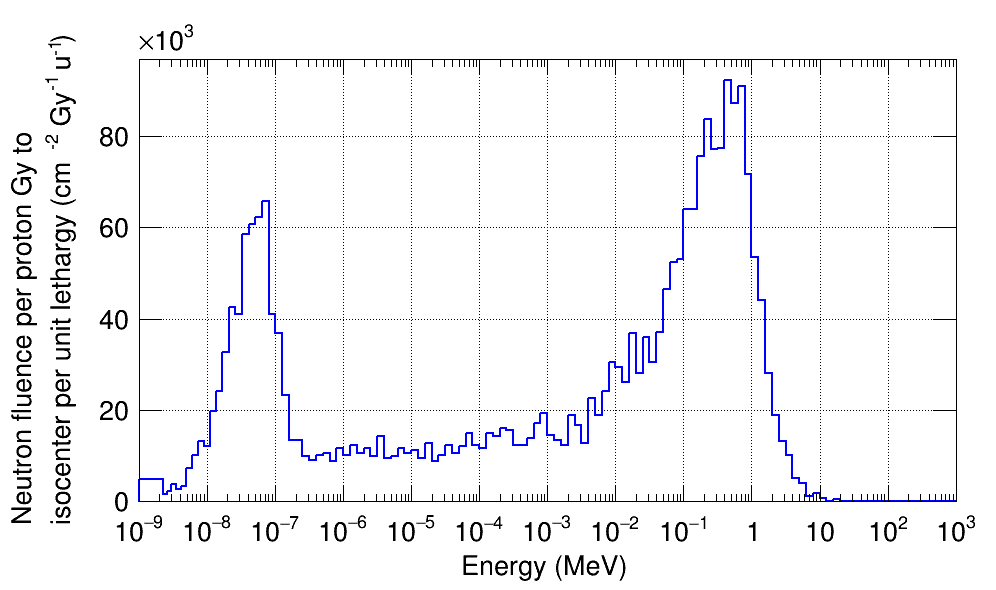}
\caption{Neutrons from the primary protons sent through nozzle}
\label {fig:2_9_b}
\end{subfigure}
\caption{Neutron fluence per unit treatment Gy incident onto the water phantom from neutrons generated in the synchrocyclotron made of stainless-steel and borated-polyethylene.}
\end{figure}

The first object that the primary protons hit is the first scatterer made of lead. Neutrons are produced as a result of direct interactions of protons with lead and move in a forward-peaked momentum distribution.

\begin{figure}[ht!]
\centering
\includegraphics[scale=0.21]{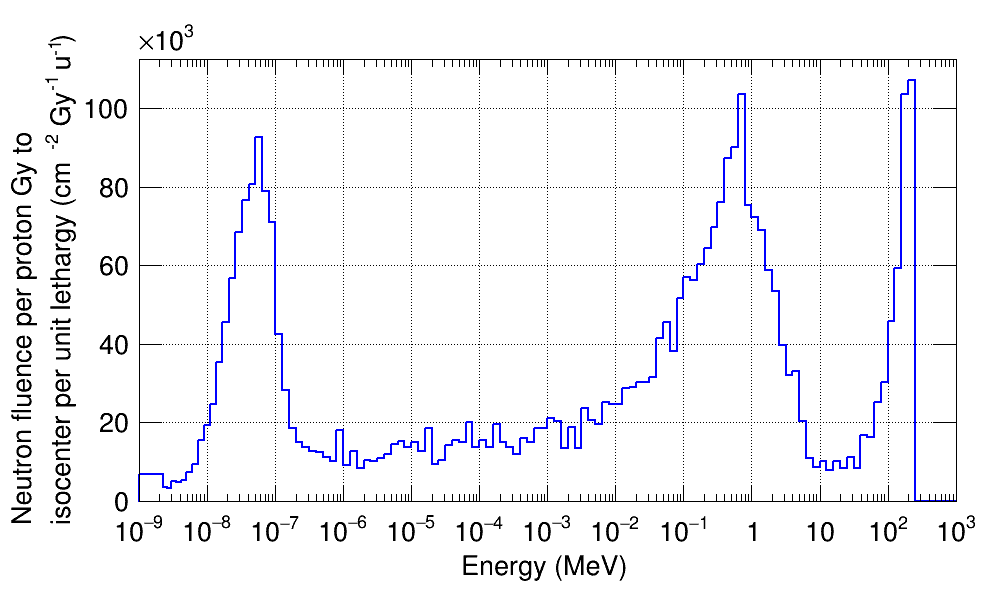}
\caption{Neutron fluence incident onto the water phantom from neutrons generated in the FS made of lead.}
\label {fig:2_10}
\end{figure}

High-energy protons interact with the different steps of the range modulation wheel containing lead. The calculated neutron spectrum at the water phantom position is similar to the neutron spectrum generated in the first scatterer.

\begin{figure}[ht!]
\begin{subfigure}{\textwidth}
\centering
\includegraphics[scale=0.21]{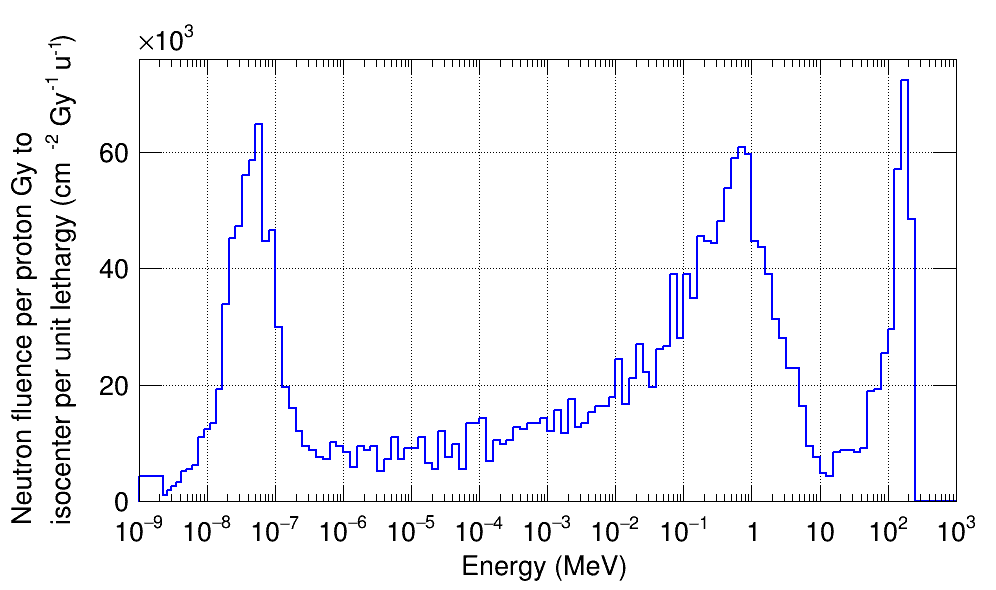}
\caption{RMW step 1}
\label {fig:2_11_a}
\end{subfigure}
\end{figure}
\begin{figure}[ht!]\ContinuedFloat
\begin{subfigure}{\textwidth}
\centering
\includegraphics[scale=0.21]{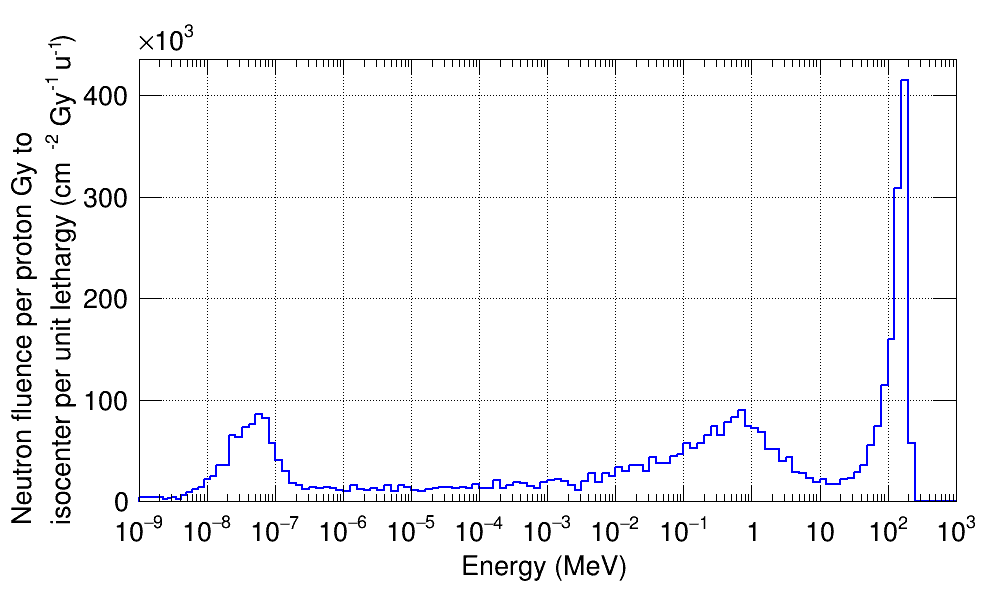}
\caption{RMW step 7}
\label {fig:2_11_b}
\end{subfigure}
\end{figure}
\begin{figure}[ht!]\ContinuedFloat
\begin{subfigure}{\textwidth}
\centering
\includegraphics[scale=0.21]{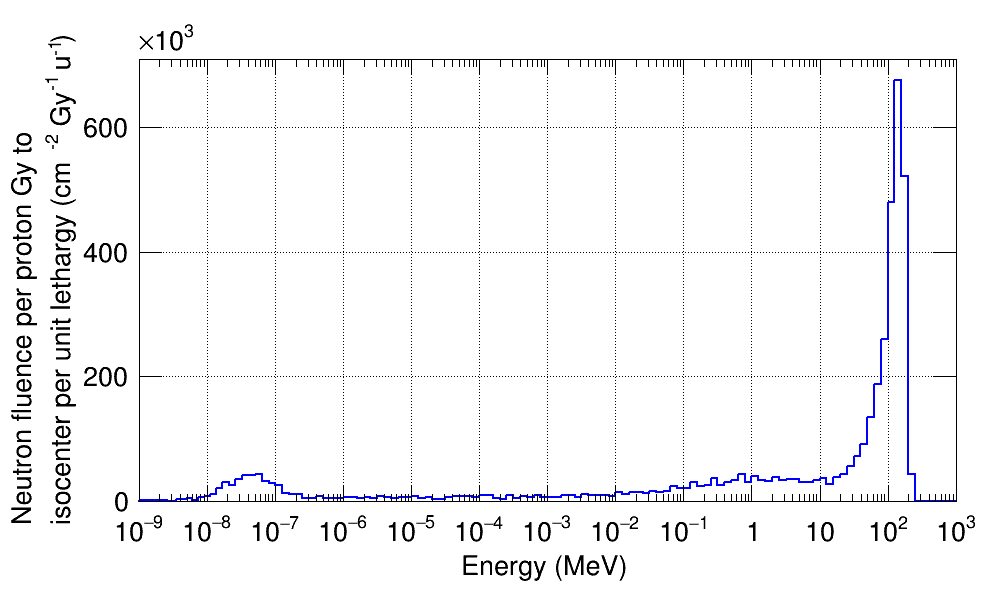}
\caption{RMW step 14}
\label {fig:2_11_c}
\end{subfigure}

\caption{Neutron fluence incident onto the water phantom from neutrons generated in the 1$^{st}$, 7$^{th}$, and 14$^{th}$ steps of the RMW}
\end{figure}

Using step 7 of the RMW increased the fast neutron peak by a factor of 5, as can be seen in Figure 11 (b). The evaporation and thermal peaks remained in the same order. The increase could be explained by the additional layer of Lexan with 1.02 cm thickness and graphite with 2.22 cm thickness. Charge exchange can happen frequently in light, symmetric nuclei due to direct reactions with the protons. This type of reaction in carbon $^{12}$C(p,n)$^{12}$N could generate fast neutrons.

Figure 11 (c) shows the neutron fluence spectrum detected at the water phantom generated by the 14$^{th}$ step of the RMW. This step contains 1.95 cm thick Lexan and 6.9 cm thick graphite. The fast neutron peak increased by a factor of 1.5 compared to the results with the 7$^{th}$ step of the RMW.

\subsubsection{Main range shifter (MRS)}
MRS was modeled as a disk with 4 cm thickness made with carbon \cite{23}⁠. A trend for the neutron fluence similar to that of the RMW steps 7 and 14 was observed here, as shown in Figure 12. The protons passing through the RMW interacted with carbon in the MRS and generated fast neutrons. The magnitude of the fast neutron component was similar to that generated in the RMW step 14 with a 6.9 cm thickness of carbon.

\begin{figure}[ht!]
\centering
\includegraphics[scale=0.21]{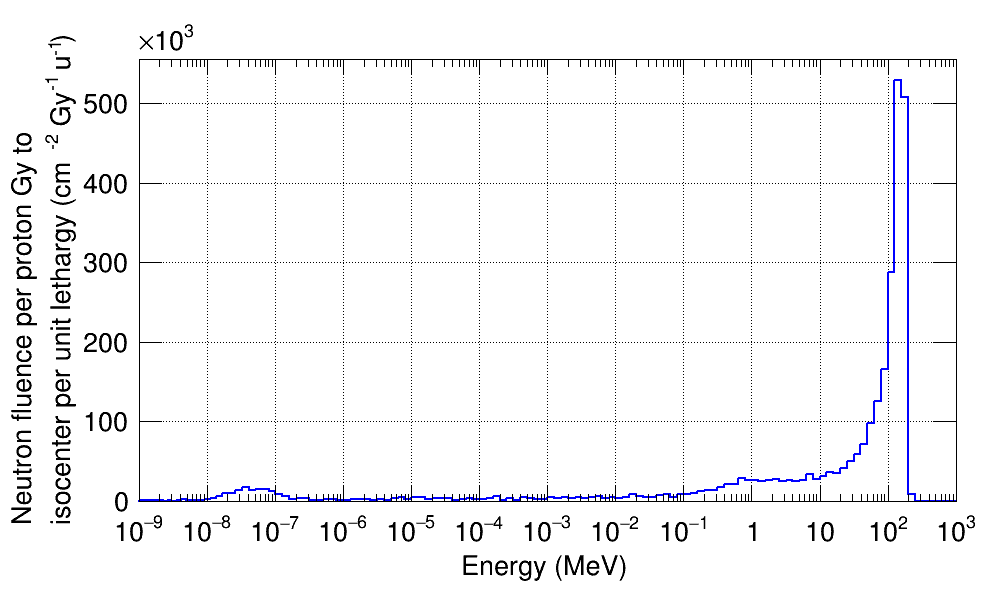}
\caption{Neutron fluence incident onto the water phantom from neutrons generated in the MRS made of carbon.}
\label {fig:2_12}
\end{figure}

\subsubsection{Second scatterer (SS)}
The SS is modeled by an outer brass ring and an inner part made of Lexan and lead. The proton beam does not mainly interact with the brass ring of the SS. Its Lexan components have a total mass of 0.52 kg. The lead has a mass of 0.06 kg. The large fast neutron peak in comparison to the evaporation peak could be associated with the carbon elements in the Lexan. Similar to the RMW and MRS, charge exchange (p,n) and knock-out (p, np) due to inelastic proton interactions could be the main processes for fast neutron production. The fluence of the water phantom is indicated in Figure 14.

\begin{figure}[ht!]
\centering
\includegraphics[scale=0.21]{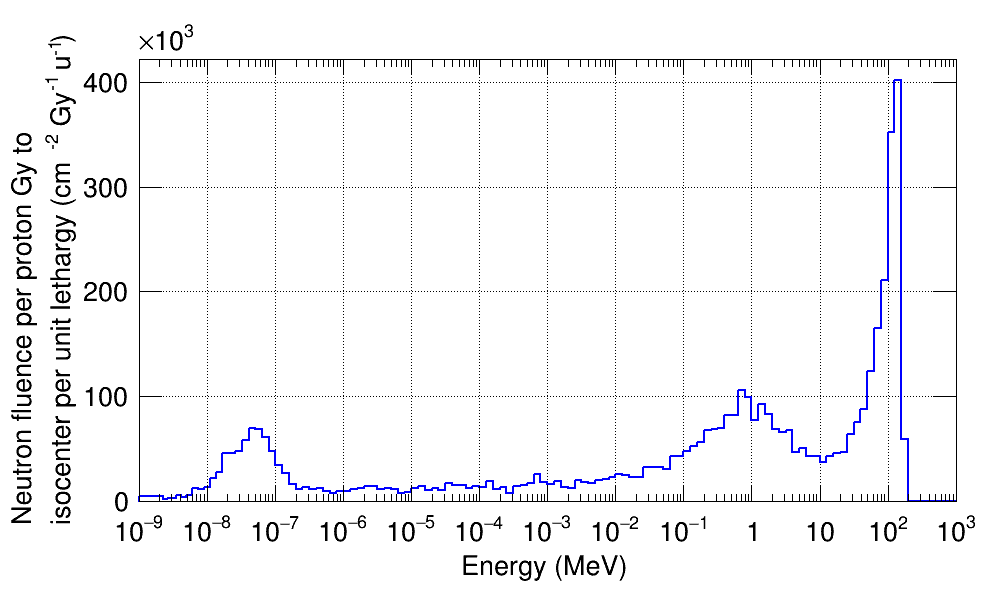}
\caption{Neutron fluence incident onto the water phantom from neutrons generated in the second scatterer made of lead, Lexan, and brass.}
\label {fig:2_13}
\end{figure}

\subsubsection{Collimation system}
The collimation system consists of a beam shielding cylinder and a collimator tube both made of stainless steel, a mixture of 74\% iron, 18\% chromium, and 8\% nickel atoms as described in the NIST material database of Geant4 \cite{43}⁠.

The main interactions in these components are initiated by the primary protons that have been degraded down to the energy region of 160 MeV – 120 MeV depending on the step used in the energy modulation wheel. Protons are tangential to the inner edges of these tubes due to the transverse scattering that is done by the first and the second scatterers. The fast neutrons are generated within the collimation system, as indicated in Figure 14.

\begin{figure}[ht!]
\begin{subfigure}{\textwidth}
\centering
\includegraphics[scale=0.21]{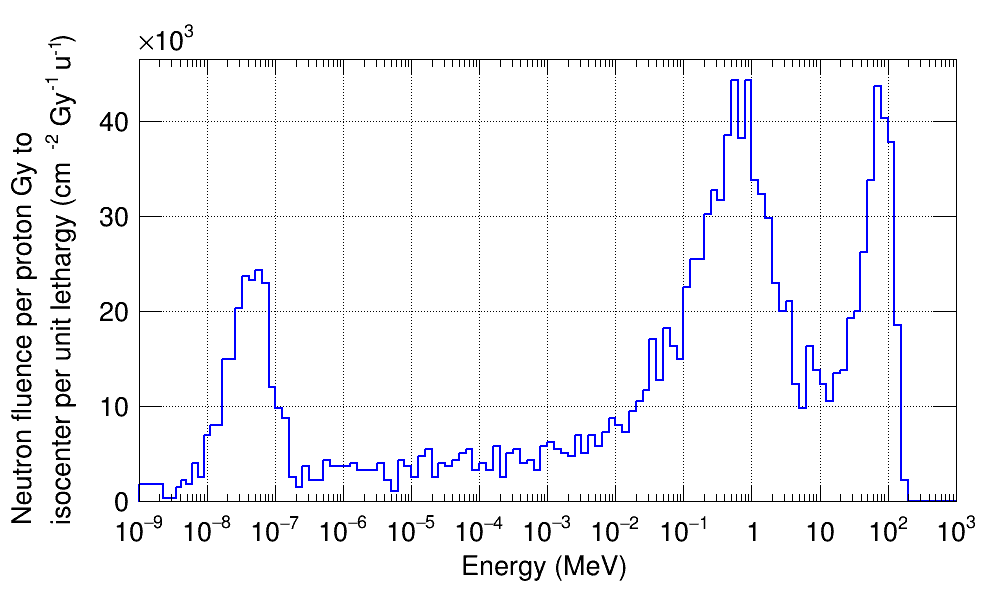}
\caption{Beam shield neutrons}
\label {fig:2_14_a}
\end{subfigure}
\end{figure}
\begin{figure}[ht!]\ContinuedFloat
\begin{subfigure}{\textwidth}
\centering
\includegraphics[scale=0.21]{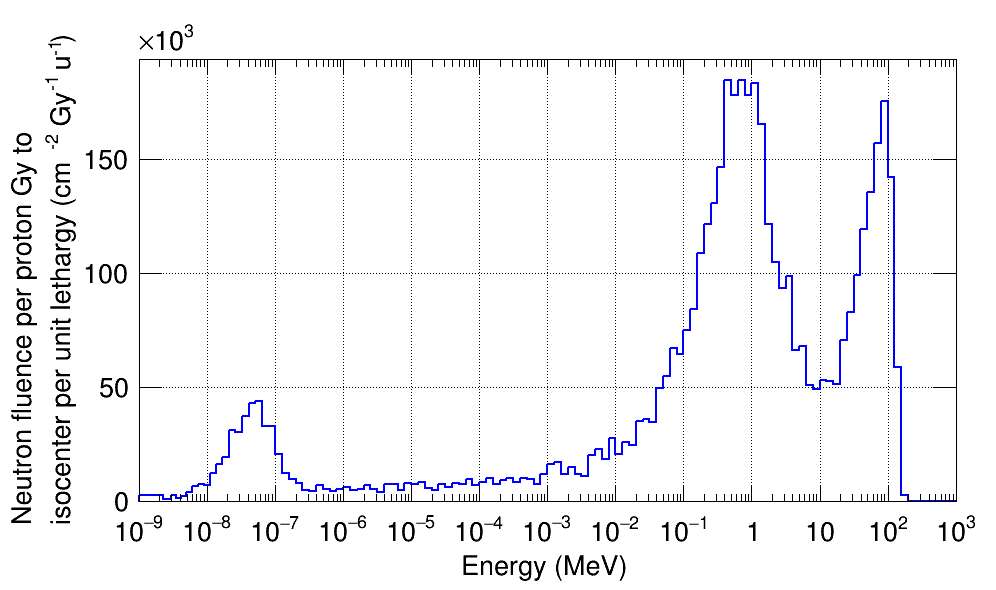}
\caption{Collimator neutrons}
\label {fig:2_14_b}
\end{subfigure}
\caption{Neutron fluence incident onto the water phantom from neutrons generated in the collimation process.}
\end{figure}

Fast neutrons mainly correspond to the charge exchange (p,n) and knock-out (p, np) reactions. The average of the cross-section for (p, nx), x representing any by-product nucleon or fragment aside from the secondary neutron, with a stainless-steel target is 0.7 barn. For $^{12}$C target,  the cross-section of (p,nx) is of the order 0.3 barn. Given this value for $^{12}$C cross-section of (p, nx), large values are observed for fast neutron peaks for the RMW and MRS that contain carbon. The beam shield causes approximately 1/10 of fast neutrons incident to the water phantom compared to the MRS, although the cross-section of fast neutron production in the beam shield is more than twice larger than that of the MRS. This could be explained by a reduced number of protons that interact with the beam shield, in addition to the peripheral position of the shield with respect to the proton beam, as compared to the central position of the MRS in the beam path. The latter fact leads the fast neutrons from the MRS to have a larger overlap with the surface of the water phantom.

The evaporation peak is due to neutrons that are emitted from excited nuclei. Fast neutron production processes, including (p,n) reactions, could form compound nuclei. A compound nucleus is “an unstable nucleus formed by the coalescence of an atomic nucleus with a captured particle” \cite{44}⁠. Such excited systems will undergo subsequent neutron evaporation, with an exception of the clustered systems such as $^{12}$C or $^{16}$O.  Clustered systems usually break up into smaller clusters such as the break-up of a compound $^{12}$C into three alphas.

In addition to proton reactions, compound systems could be formed by neutron nonelastic reactions that can lead to further neutron evaporation. Neutrons are thermalized due to elastic reactions with light targets. Hydrogen provides a high thermalizing ratio per interaction. Thermal neutrons are captured by nuclei. Neutron capture excites the nuclei and causes gamma radiation. Due to radiative capture, neutron fields are mixed with gamma. According to the measurements performed with a WENDI-2 chamber, the absorbed dose equivalent from gamma is less than 3\% of that from neutrons.

\subsubsection{Brass aperture (BA)}
The neutron spectrum incident on the water phantom (WP) from those neutrons that have been initially generated in the BA shows a low thermal-to-evaporation peak ratio. The integral of the neutron fluence incident onto the WP from the BA is significantly larger than the other components. This is due to the small distance between the BA and the WP. The forward-peaked fast neutrons have a large probability to hit the WP surface because of the short distance.

Also, the low thermal-to-evaporation peak ratio in the spectrum shown in Figure 15 could be explained by the small amount of material between the BA to WP which reduces the probability for the evaporation neutrons to undergo elastic reactions with light nuclei and thermalize.

Brass is made of 67.4\% copper atoms, 32.1\% zinc atoms, and less than 0.5\% mixture of iron, lead, and tin atoms. The primary proton incident on the brass aperture is mostly in this energy range. These protons have been degraded in energy by the first scatterer, the range modulation wheel, and the second scatterer.

Fast neutron production rate in brass per high-energy protons in the range of 120 to 160 MeV impinging on the brass aperture is expected to be in the same order of magnitude as that in $^{12}$C which was discussed in the subsection on the RMW. The elements in the brass aperture and the graphite in the RMW have a $\sigma$ of about 0.4 and 0.7 barns for nonelastic (p, nx) reactions, respectively.

\begin{figure}[ht!]
\centering
\includegraphics[scale=0.21]{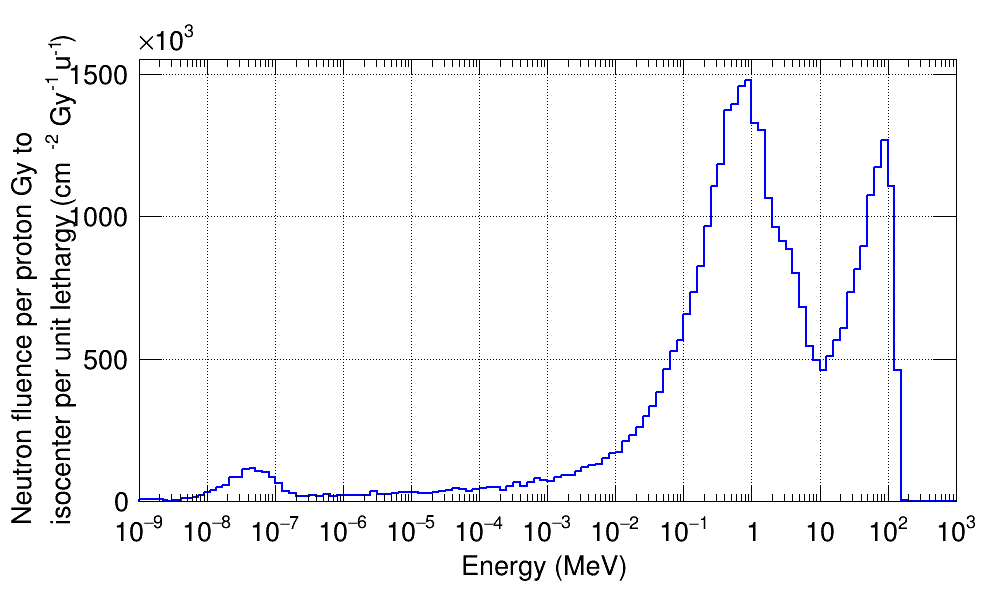}
\caption{Neutron fluence incident onto the water phantom from neutrons generated in the brass aperture.}
\label {fig:2_15}
\end{figure}

The nonelastic (n,nx) cross-section values for various neutron energies in the MeV range, could be used to explain the evaporation of neutrons produced in the brass aperture. Values of the cross-sections for (n,nx) reactions in brass between incident neutron energies of 1 MeV to 10 MeV are comparable in magnitude to the cross-section of (p,n) reactions in $^{12}$C with $\sigma$=0.7 barn. It was observed that (p,n) reaction in $^{12}$C leads to a large fast neutron fluence incident on the water phantom. With a similar analogy, we might conclude a large probability for the evaporation process based on the  $\sigma$ values for the (n, nx) process in brass. That could explain the large evaporation peak observed in Figure 15.

\subsubsection{Ceiling and floor}
The ceiling and floor were simulated by 15 cm and 20 cm thick concrete respectively, with a 150 cm vertical distance from the isocenter each.

About half of the atoms in concrete are oxygen. The most probable reactions that could happen on the floor would be from incident neutrons with various energies, lower than the maximum primary proton energy. Neutrons that reach the floor have interacted several times elastically or inelastically and have attenuated in energy. Based on this fact, the absence of the fast neutron peak in figures 16 (a) and (b) could be explained.

Neutrons with energies below 1 MeV have a large cross-section of interaction with hydrogen nuclei. Hydrogen forms 30\% of the total nuclei in concrete. The attenuation ratio of neutron energy by elastic interactions with hydrogen per collision is 0.5, on average. The thermal neutron peaks, as shown in Figure 16, were close to the evaporation peak in magnitude. Also, a relatively large continuum of neutrons between thermal and evaporation energies was produced by the simulations.

\begin{figure}[ht!]

\begin{subfigure}{\textwidth}
\centering
\includegraphics[scale=0.21]{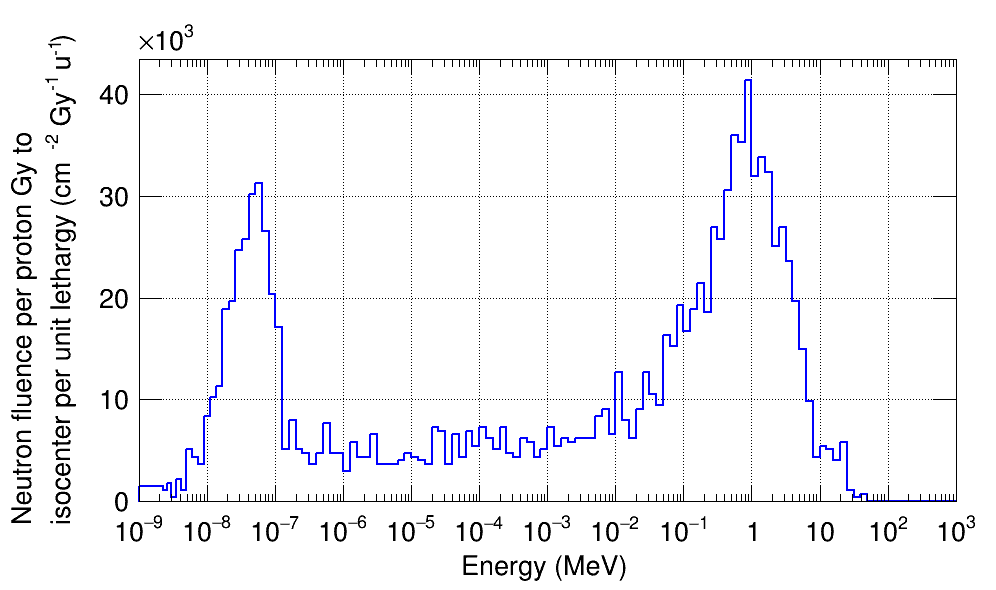}
\caption{Ceiling neutrons}
\label {fig:2_16_a}
\end{subfigure}
\end{figure}
\begin{figure}[ht!]\ContinuedFloat
\begin{subfigure}{\textwidth}
\centering
\includegraphics[scale=0.21]{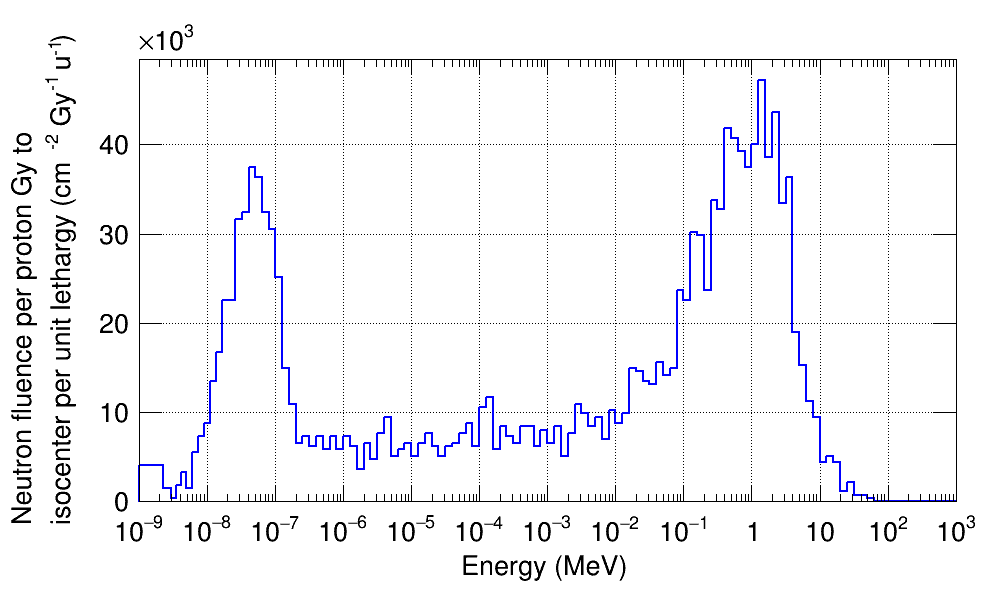}
\caption{Floor neutrons}
\label {fig:2_16_b}
\end{subfigure}
\caption{Neutron fluence incident onto the water phantom from neutrons generated in the treatment room.}
\end{figure}

\subsubsection{Internal neutrons}
The calculated fluence for the internal neutrons in the water phantom has been done by recording the energy of each neutron once produced inside the water phantom. Similar to the fluence diagrams from other components, the resulting internal neutron histogram has been divided by the average cross-sectional area of the water phantom.

According to Figure 17, no neutrons were generated in the thermal region. This is due to the fact that the neutrons are recorded at the initial step of their products before being thermalized. Also, the magnitude of the fast and evaporation neutrons are close, with the fast neutron peak being slightly larger.

\begin{figure}[ht!]
\centering
\includegraphics[scale=0.21]{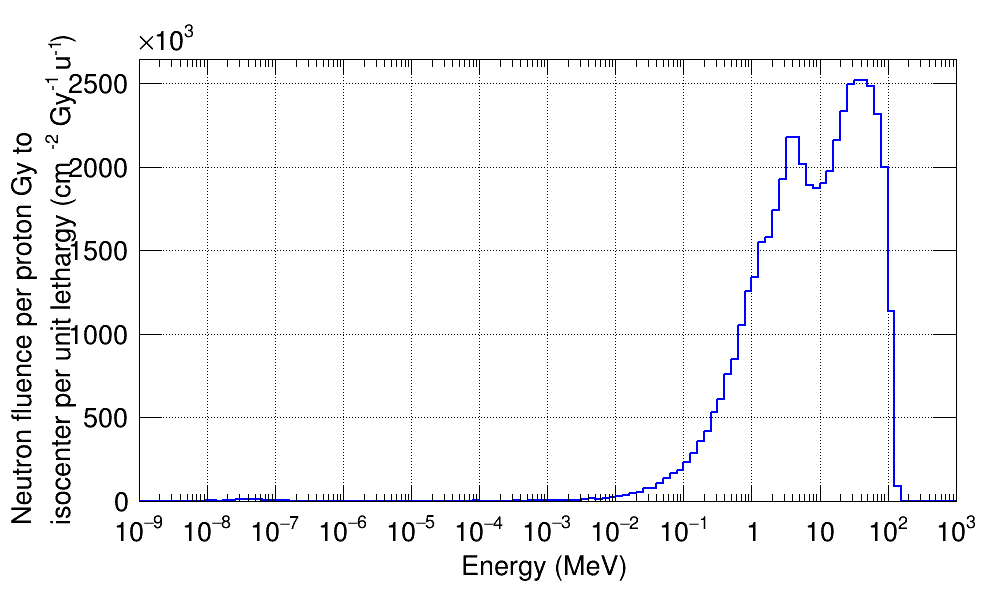}
\caption{Internal neutron fluence from neutrons generated in the water phantom.}
\label {fig:2_17}
\end{figure}

\subsubsection{Comparison of neutron fluence and neutron fluence per unit mass of the nozzle items}
Figure 18 shows the neutron fluence spectra per unit treatment Gy to the isocenter per unit lethargy for all the nozzle components. According to Figure 18, the largest contribution to the external fluence received at the water phantom comes from the brass aperture. Neutron fluence from other items in the nozzle is about an order of magnitude less than the brass aperture's contribution. The abbreviations in the key in Figures 18 and 2.19 read as: FS= first scatterer, RMW= range modulation wheel, MRS= main range shifter, SS= second scatterer, SHLD= beam shield, CLMT= beam collimator, BRSAP= brass aperture, CMPNS= Lucite compensator.

\begin{figure}[ht!]
\centering
\includegraphics[scale=0.21]{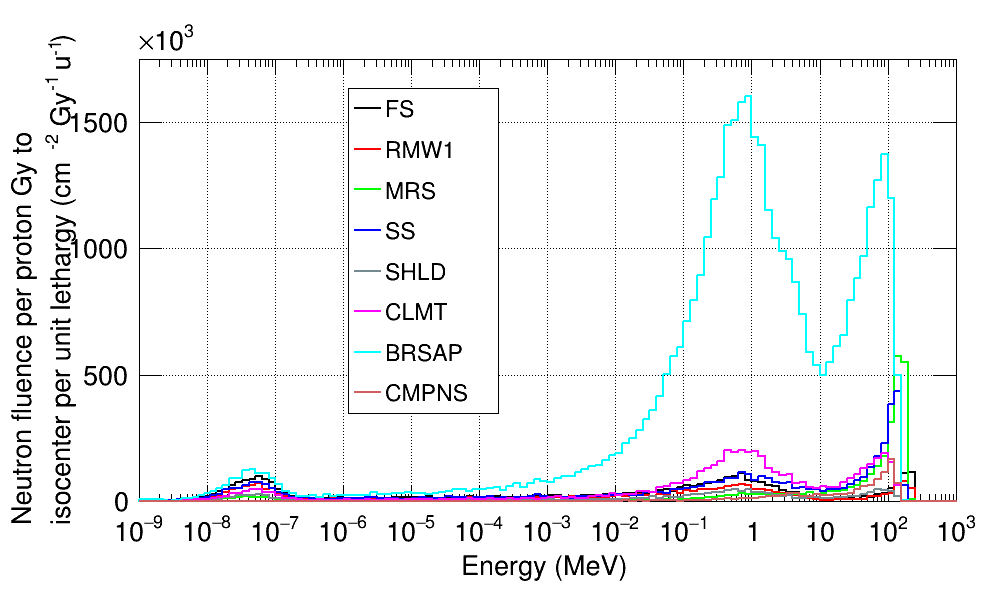}
\caption{Neutron fluence spectra incident onto the water phantom from neutrons generated in the nozzle's items. The contribution from the brass aperture dominates.}
\label {fig:2_18}
\end{figure}

Figure 19 shows the neutron fluence spectra per unit treatment Gy to the isocenter per unit lethargy for all the nozzle components per their unit mass. According to Figure 19, the largest contribution to the external fluence received at the water phantom per unit mass of the items comes from the first scatterer and the first step of the range modulation wheel. This fact is due to the lead component in these items that contributes largely to fast neutron production. Neutron fluence per unit mass from other items in the nozzle is about an order of magnitude less than the brass aperture contribution.

\begin{figure}[ht!]
\centering
\includegraphics[scale=0.21]{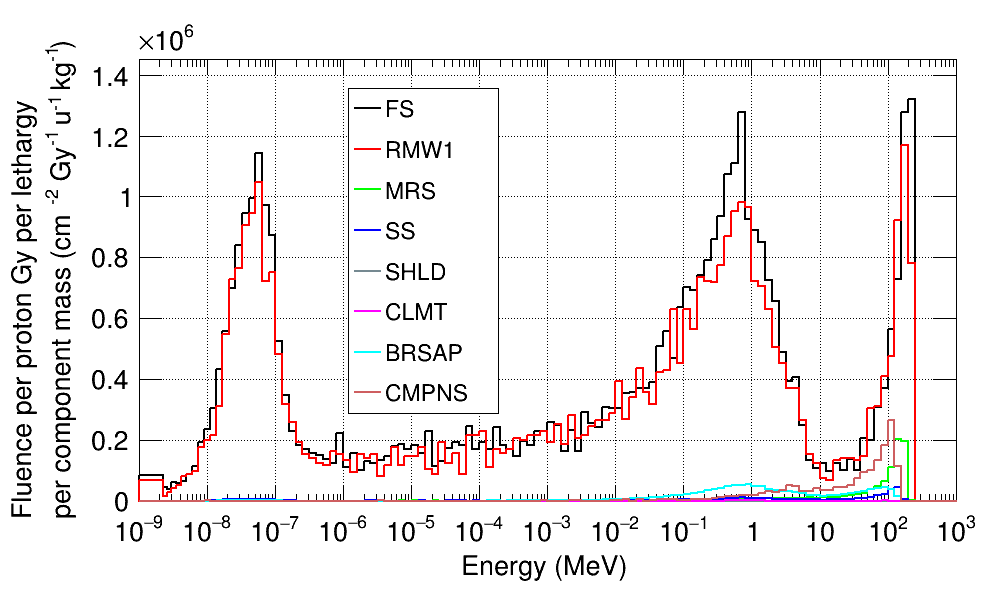}
\caption{Neutron fluence spectra incident onto the water phantom from neutrons generated in the nozzle's items per unit mass of the corresponding item. The first scatterer and the range modulation wheel dominate.}
\label {fig:2_19}
\end{figure}

\subsection{Neutron profiles inside the treatment room with and without the water phantom}
For this section, the profiles of the neutron dose-equivalent per treatment Gy have been plotted and the general trends of the neutron dose behaviors at various positions inside the treatment room have been obtained. The simulations for calculating the profiles in Figures 22 and 2.23 were done with a water phantom with dimensions 35$\times$35$\times$35 cm$^3$ placed at the isocenter. The first RMW step was used and the brass aperture opening was set to 20$\times$20 cm$^2$.

\subsubsection{Neutron dose profile perpendicular to the beam direction}
Figure 20 (a) plots the profiles for the absorbed neutron dose-equivalent per treatment Gy to the isocenter for different heights measured from the isocenter. The profiles show the behavior of H*/D as a function of distance along the x-axis, which is the axis parallel to the synchrocyclotron cylinder. As can be seen from Figure 20 (a), an approximately symmetric behavior was observed for H*/D. As seen for the heights 80 cm and 120 cm, some deviations from the symmetric behavior could be due to the asymmetry inside the room. The entrance of the maze into the treatment room introduces a source of asymmetry that might affect the neutron dose-equivalent profiles. The maximum value observed for H*/D was 9.2 mSv/Gy, which was due to the neutrons counted at the isocenter. A comparison between the two cases where a water phantom was placed at the isocenter and where it was not there has been provided in Figure 20 (b).

\begin{figure}[ht!]
\begin{subfigure}{\textwidth}
\centering
\includegraphics[scale=0.21]{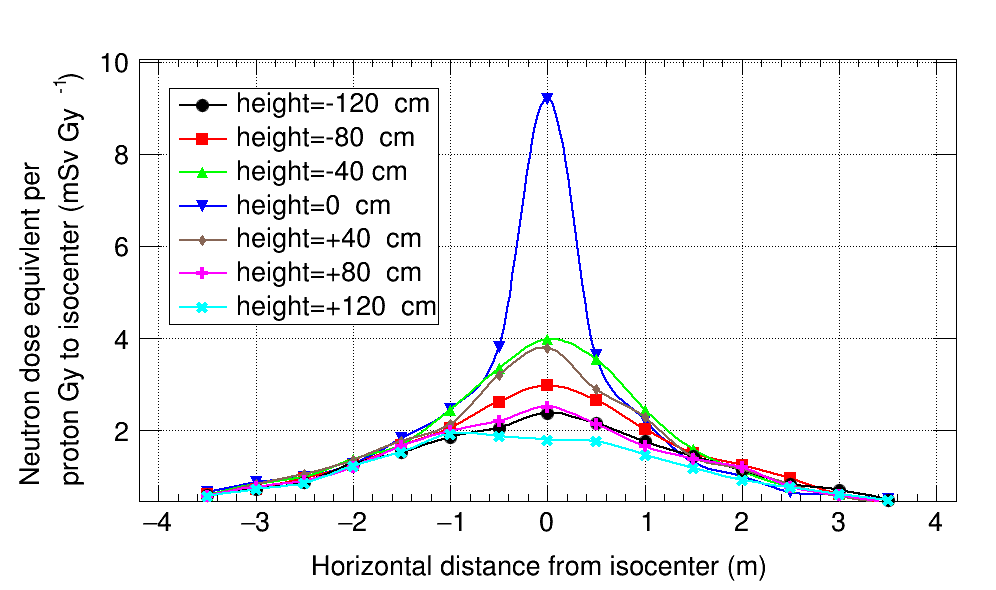}
\caption{At various heights, with water phantom.}
\label {fig:2_20_a}
\end{subfigure}
\end{figure}
\begin{figure}[ht!]\ContinuedFloat
\begin{subfigure}{\textwidth}
\centering
\includegraphics[scale=0.21]{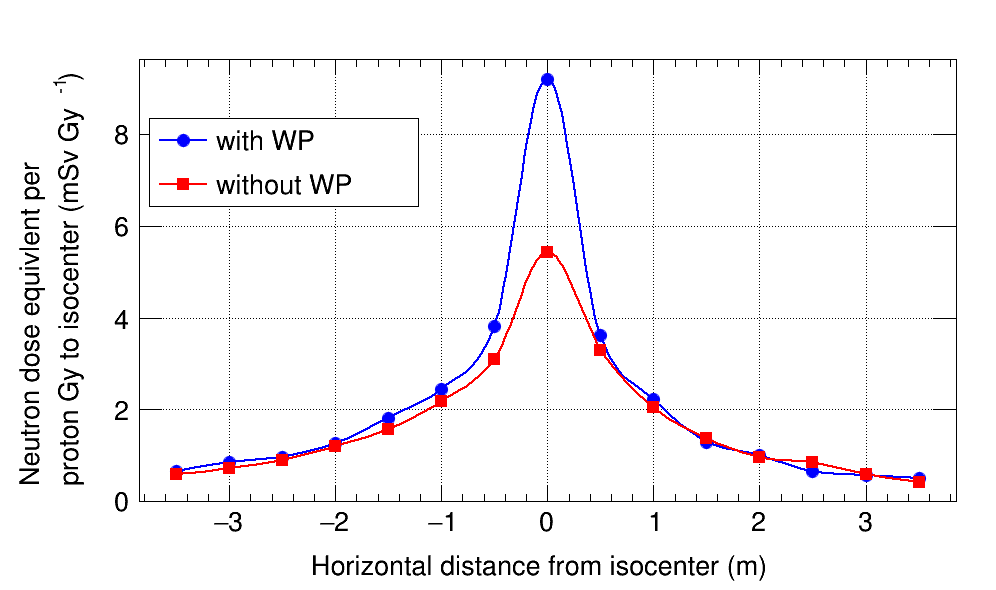}
\caption{With and without water phantom.}
\label {fig:2_20_b}
\end{subfigure}
\caption{Neutron dose-equivalent profiles per treatment proton dose at the isocenter in mSv/Gy unit. The values have been plotted as a function of transverse distance from the isocenter across the x-axis for various heights. A water phantom with dimensions 35$\times$35$\times$35 cm$^3$ has been placed at the isocenter for the results shown in figure (a). The profiles in figure (b) correspond to the H*/D values across the x-axis at height y=0 from the isocenter with a water phantom (blue) and without a water phantom (red). The first RMW step was used and the brass aperture opening was set to 20$\times$20 cm$^2$.}
\end{figure}

\subsubsection{Neutron dose profiles along the beam direction}
Figure 21 plots the profiles for the absorbed neutron dose-equivalent per treatment Gy to the isocenter along the z-axis at a height 50 cm below the isocenter and parallel to the proton beam. The blue curve corresponds to a simulation with a water phantom placed at the isocenter and the red curve corresponds to a simulation without a water phantom. As can be seen from Figure 21, the general trend of H*/D was a decrease in the neutron dose with increasing distance from the nozzle. An exception to this trend was found at the isocenter, where the water phantom was positioned. The internal neutrons generated inside the water phantom shifted the profiles upward at the region around z=0. In Figure 21, the point z=0 indicates the isocenter, and the primary protons are generated at the point z=2 m toward the negative z-direction. The calculated value of H*/D at the isocenter was 4 mSv/Gy when the water phantom front side was positioned and 3.5 mSv/Gy when it was removed. The largest difference between the two cases was observed at z=0.5 m, close to the backside of the water phantom. At that position, 2.5 times larger H*/D was scored for the configuration with the water phantom. At z=2 m, the extraction point where the primaries are generated, the value of H*/D for both cases was about 19 mSv/Gy. Finally, closer to the front wall at z=-6.5 m, a slight increase in the H*/D was observed for the case without the water phantom. That could be associated with the evaporation neutrons that back-scatter from the front wall due to the high energy protons that reach the wall because of the absence of blocking by the water phantom.

\begin{figure}[ht!]
\centering
\includegraphics[scale=0.21]{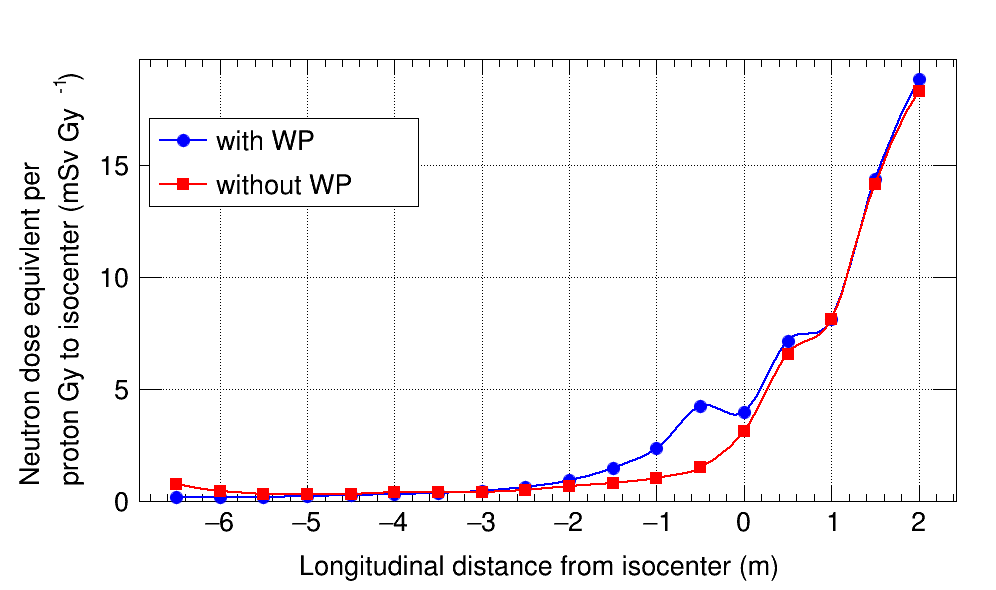}
\caption{Neutron dose-equivalent profile per treatment proton dose at the isocenter in mSv/Gy unit. The values have been plotted as a function of distance along the beam direction on an axis 50 cm below the isocenter. Isocenter is at z=0, positive z is toward the nozzle, and the proton beam is directed toward the negative z-direction. A water phantom with dimensions 35$\times$35$\times$35 cm$^3$ has been placed at the isocenter. The first RMW step was used and the brass aperture opening was set to 20$\times$20 cm$^2$.}
\label {fig:2_21}
\end{figure}

\section{Conclusions}
A study of neutron production in the scattering Mevion proton-therapy system is performed. The influence of several parameters on neutron production in the scattering Mevion machine was considered and analyzed. The Geant4 code was calibrated using previously reported data. The results for the total fluence and total neutron dose-equivalent matched the extended-range Bonner sphere (ERBS) measurements within 4.6\% and 0.25\%, respectively. In addition, the neutron dose-equivalent calculations at 5 transverse points with comparable beam and nozzle parameters to a previous TOPAS study were fit with a difference of an order of 10\%.

The simulations with Geant4, however, did not agree with the WENDI-2 measurements in the same study that provided the TOPAS results. Moreover, other reported values of H*/D exist in the literature that my data has not agreed with. These differences could be associated with the various types of machines, the different proton source average energies, energy spreads, and spot sizes, uncertainties regarding the proton extraction ratios in the accelerators, the geometrical uncertainties in the items' dimensions and variations in the material densities, and to the uncertainties in the cross-section libraries.

The analysis of neutron production in the scattering machine consists of showing several diagrams of the neutron fluence spectra per treatment Gy to the isocenter received from various items in the nozzle and those in the room onto the water phantom, which models a human body. Data on the percentages of thermal, intermediate, evaporation, and fast neutron production were provided for each contributing item to the neutron dose. The neutron fluence spectra per unit mass of each item are shown. Correspondingly, the largest neutron production per mass of the objects is from the lead in the first scatterer and in the RMW.

The pie charts of the total H*/D from different nozzle components and room items are obtained and plotted. An analysis is done by changing the RMW step at a fixed field size and changing the field size at a fixed RMW step. The corresponding effects on the values of H*/D are observed and studied. It is concluded from the pie charts that about 60\% of the neutron H*/D is caused due to the internal neutrons when the 1$^{st}$ RMW step is used. This ratio decreased to 50\% for 7$^{th}$ RMW step. Using the 14$^{th}$ step of the RMW further decreases this ratio to about 25\%.

A decrease of H*/D from the neutrons generated in the brass aperture is observed as a result of rotating the RMW from the first step upward. For the 1$^{st}$ RMW step the contribution of the brass aperture is 2.7 mSv/Gy, for the 7$^{th}$ RMW step it is 1.4 mSv/Gy, and for the 14$^{th}$ RMW step it is 0.28 mSv/Gy. In this analysis, the RMW output beam energy is lowered from 227 MeV to 165 MeV, while keeping the brass aperture opening constant with a 10 cm side length. As a result of this shift in the RMW output beam energy, the absorbed neutron H*/D to the water phantom decreases by about a factor of 10.

The profiles of the total, i.e. integrated, H*/D values as functions of position inside the treatment room is obtained. A general decreasing trend of H*/D is observed from the proton extraction point toward the isocenter and beyond. The water phantom is found to cause an increase in the H*/D profile around the isocenter. This fact is discussed to be due to the internal neutrons that move outward from the water phantom. In conclusion, a set of the expected values is presented for the absorbed neutron dose-equivalent per unit treatment Gy to the isocenter for the scattering Mevion proton machine at various positions, and as a function of various parameters. The H*/D values could be used to obtain a better understanding of the difference between the neutron production in the scattering Mevion S250 system and the scanning Mevion S250i system that was studied in Chapter 3.

\printbibliography[heading=bibintoc, title={References}]

@article{3,
author=			"R. Baskar and K. A. Lee and R. Yeo and K. W. Yeoh",
title=			"{Cancer and Radiation Therapy: Current Advances and Future Directions}",
journal=		"{Int. J. Med. Sci.}",
volume=			"{9}",
number=			"{3}",
pages=			"{pp 193-199}",
year=			"{(2012)}",
}

@article{9,
author=			"Tianyu Zhao and Baozhou Sun and Kevin Grantham and Leith Rankine and Bin Cai and Sreekrishna M. Goddu and Lakshmi Santanam and Nels Knutson and Tiezhi Zhang and Michael Reilly and Beth Bottani and Jeffrey Bradley and Sasa Mutic and Eric E. Klein",
title=			"{Commissioning and Initial Experience with The First Clinical Gantry-Mounted Proton Therapy System}",
journal=		"{J. Appl. Clin. Med. Phys.}",
volume=			"{17}",
pages=			"{pp 24-40}",
year=			"{(2016)}",
doi=			"{10.1120/jacmp.v17i2.5868}",
}

@article{12,
author=			"M. Schippers",
title=			"{Proton Accelerators}",
journal=		"{CRC press, Proton Therapy Physics}",
year=			"{(2012)}",
}

@article{16,
author=			"B. Jones and T. S. A. Underwood and C. Timlin and R. G. Dale",
title=			"{Fast Neutron Relative Biological Effects and Implications for Charged Particle Therapy}",
year=			"{(2011)}",
doi=			"{10.1259/bjr/67509851}",
}

@article{23,
author=			"Milad Baradaran-Ghahfarokhi and Francisco Reynoso and Baozhou Sun and Arash Darafsheh and Michael T. Prusator and Sasa Mutic and Tianyu Zhao",
title=			"{A Monte Carlo-based Analytic Model of Neutron Dose Equivalent for a Mevion Gantry-Mounted Passively Scattered Proton System for Craniospinal Irradiation}",
journal=		"{Med. Phys.}",
volume=			"{47}",
pages=			"{pp 4509–4521}",
year=			"{(2020)}",
doi=			"{10.1002/mp.14299}",
}

@article{24,
author=			"Rebecca M. Howell and Eric A. Burgett and Daniel Isaacs and Samantha G. Price Hedrick and Michael P. Reilly and Leith J. Rankine and Kevin K. Grantham and Stephanie Perkins and Eric E. Klein",
title=			"{Measured Neutron Spectra and Dose Equivalents from a Mevion Single-Room, Passively Scattered Proton System Used for Craniospinal Irradiation}",
journal=		"{Int. J. Radiat. Oncol. Biol. Phys.}",
volume=			"{95}",
pages=			"{pp 249–257}",
year=			"{(2016)}",
doi=			"{10.1016/j.ijrobp.2015.12.356}",
}

@article{39,
title=			"{ENDF Neutron Cross-Sections}",
url=			"https://www-nds.iaea.org/exfor/endf.htm",
urldate=		"2021-08-02",
}

@article{43,
url=			"https://geant4-userdoc.web.cern.ch/UsersGuides/ForApplicationDeveloper/html/Appendix/materialNames.html",
title=			"{NIST Materials}",
urldate=		"2021-08-02",
}

@article{44,
url=			"https://www.merriam-webster.com/dictionary/compound nucleus",
title=			"{Merriam Webster}",
urldate=		"2021-08-02",
}

\end{document}